\documentclass{jfm}

\usepackage{CJK}

\usepackage{ulem}
\usepackage[T1]{fontenc}
\usepackage{calligra}

\usepackage{graphicx}% Include figure files
\usepackage{dcolumn}% Align table columns on decimal point
\usepackage{bm}% bold math

\usepackage{times}
\usepackage{verbatim}
\usepackage{graphicx}
\usepackage{graphics,epsfig}

\usepackage{theorem}
\usepackage{makeidx}
\usepackage{amsmath}
\usepackage{epic}
\usepackage{amscd}

\usepackage{bbm}
\usepackage{xy}
\usepackage{amssymb}
\usepackage{epsfig}

\usepackage{cancel}
\usepackage{graphicx}
\usepackage{natbib}

\ifCUPmtlplainloaded \else
  \checkfont{eurm10}
  \iffontfound
    \IfFileExists{upmath.sty}
      {\typeout{^^JFound AMS Euler Roman fonts on the system,
                   using the 'upmath' package.^^J}%
       \usepackage{upmath}}
      {\typeout{^^JFound AMS Euler Roman fonts on the system, but you
                   dont seem to have the}%
       \typeout{'upmath' package installed. JFM.cls can take advantage
                 of these fonts,^^Jif you use 'upmath' package.^^J}%
      }
  \else
  \fi
\fi

\ifCUPmtlplainloaded \else
  \checkfont{msam10}
  \iffontfound
    \IfFileExists{amssymb.sty}
      {\typeout{^^JFound AMS Symbol fonts on the system, using the
                'amssymb' package.^^J}%
       \usepackage{amssymb}%
       \let\le=\leqslant  
       \let\ge=\geqslant  
      }{}
  \fi
\fi

\ifCUPmtlplainloaded \else
  \IfFileExists{amsbsy.sty}
    {\typeout{^^JFound the 'amsbsy' package on the system, using it.^^J}%
     \usepackage{amsbsy}}
    {}
\fi

\newsavebox{\astrutbox}
\sbox{\astrutbox}{\rule[-5pt]{0pt}{20pt}}

\title[
Chiroid absolute equilibria and turbulence
]{
Purely helical
absolute equilibria and %one-chiral-sector-dominated turbulent states
chirality of (magneto)fluid turbulence
}

\author[J.-Z. Zhu
\& W. Yang and G.-Y. Zhu
]%
{Jian-Zhou Zhu$^{
}$%
, Weihong Yang$^{2
}$ %
and Guang-Yu Zhu$^{1,
3}$
}

\affiliation{$^1$WCI Center for Fusion Theory,
National Fusion Research Institute, 169-148 Gwahak-ro, Daejeon, Korea\\[\affilskip]
$^2$Department of Modern Physics, University of Science and Technology of China, 230026 Hefei, China\\[\affilskip]
$^3$Life and Chinese Medicine Study Center, 47 Bayi Cun, 366025 Yong'an, Fujian, China}

\pubyear{2010}
\volume{650}
\pagerange{119--126}
\date{?; revised ?; accepted ?. - To be entered by editorial office}
\begin{document}

\maketitle

\begin{abstract}
Purely helical absolute equilibria of incompressible neutral fluids and plasmas (electron, single-fluid and two-fluid magnetohydrodyanmics) are systematically studied with the help of helical (wave) representation and truncation, for genericities and specificities about helicity. A unique chirality selection and amplification mechanism and relevant insights, such as the one-chiral-sector-dominated states, among others,
about (magneto)fluid turbulence follow. \end{abstract}

\section{Background, technique, and basic ideas}\label{sec:hae}
Helical modes are basic in electromagneto- and hydrodynamics \citep[see, {\it e.g.},][]{Moses71}. They are left- or right-handed, signaturing \textit{chirality}\footnote{This notion is widely used in chemistry, physics and (origin of) life sciences and was called \textit{dissymmetry}, which is still occasionally used, before \citet{Kelvin1904} and various attempts have been made to mathematically quantify it \citep[see, {\it e.g.}, the review by][]{Petitjean03}.} which may be quantified by helicity and its relevant derivatives, such as the relative helicity% \citep{k73}
, important for the statistical dynamics.

\subsection{Helical turbulence and absolute equilibrium}\label{sec:HTaAE}
Recognizing the importance of helicity in hydrodynamic turbulence is relatively new, though Helmholtz-Kelvin theorem is old \citep[][]{Moffatt08}.
Indeed,
in a communication with C.-C. Lin in 1945% \citep[the letter is reproduced in][]{EyinkSreenivasan06}
, L. Onsager noticed that the coefficients of the Fourier modes of hydrodynamic velocity field are ```momentoids' in the sense of Boltzmann, and the theorem of equipartition would apply if their number were finite. Since this is not the case, we get a `violet catastrophe' instead.''\footnote{This remarkable comment adds more to Onsager's ignored legacy on hydrodynamics than that exposed by \citet[][private communication in 2008]{EyinkSreenivasan06} who reproduced the letter.}
Statistical absolute equilibrium (AE) energy equipartition among each Fourier modes was later explicitly formulated by T.-D. \citet{Lee1952}
for both pure hydrodynamics (HD) and magnetohydrodynamics (MHD). 
Neither Onsager nor Lee (who, interestingly, as well-known, however soon suggested with Yang in 1956 the chiral ``world'' --- parity is violated in the weak interactions!) considered the invariance of helicity\footnote{Later, \citet{Betchov61} first tried to explore %reported, with minor mistakes \citep{FrischETC75},
invariant helicity's role in turbulence, contrasting a box of nails to screws with reflexion asymmetry.} which makes the flow field lose mirror symmetry and which can also be involved in the generalized equipartitions %, as were updated by
\citep{k73,FrischETC75}. Now, tremendous progresses with helical representation/decomposition %k73,W92,BrandenburgDoblerSubramanian02
have been made \citep[see, e.g.,][and references therein]{YangWu10}:
Nature of the triadic interactions can be exposed more clearly \citep{W92} and be exploited to understand better the fluctuations, for instance, of electron MHD (EMHD) and Hall MHD \citep{GaltierBhattacharjee03,GaltierHMHD06}. And, spectral dynamics can be diagnosed in a finer way \citep[][see more detailed discussions in \S\ref{sec:HD}]{CCE03}. In particular, for the pure-helical-mode subsystem with only modes of one chiral sector of Navier-Stokes equations, one can expect %\citep[see, {\it e.g.},][]{W92}
a dual cascade picture. Indeed, we will see that \citet{k73}'s helical spectra can be refined to allow a negative temperature state %, as in the two-dimensional (2D) case \citep{k67}, with energy condensation at large scales
to support the dual cascades. On the other hand, approaching nearly maximally helical, {\it i.e.}, one-chiral-sector-dominated states (OCSDSs) with severe \textit{chiral symmetry} breaking, {\it i.e.}, imbalance of positive and negative helicity, along scales have been explicitly demonstrated \citep[{\it e.g.},][see more detailed discussions in \S\ref{sec:sfMHD}]{MeneguzziPRL81,BrandenburgDoblerSubramanian02} but want a corresponding theoretical understanding, as we will offer.
Recently, \citet{MeyrandGaltierPRL12} studied Hall MHD new chirality symmetry breaking in the sense of domination by whistler or ion-cyclotron waves defined by the linear wave dispersion relation, which is different to the chirality signature coming from the helical representation used in this paper; see \S\ref{sec:tfMHD} for more remarks.

%\subsection{Helical AE}\label{sec:hae}
The equilibrium-statistical-mechanics approach to investigate turbulence had been somewhat esoteric, but
\citet[][hereafter K67 and K73]{k67,k73} established in a more explicit and complete way the AE for both 2D and 3D incompressible HD. Fourier modes beyond $[k_{min}, k_{max}]$ being discarded (Galerkin truncation), certain rugged quadratic invariants --- for solutions regular enough to bear no dissipative anomaly \citep[see, {\it e.g.},][]{EyinkEulerPhD08} --- such as the kinetic energy ($\mathcal{E}_K$) and enstrophy (for 2D) or helicity ($\mathcal{H}_K$ for 3D), are still conserved.
With the constraints of these rugged invariants, Kraichnan obtained the respective energy spectral densities for 2D and 3D:
$$\overline{U}_K(\bm{k})=1/(\alpha+\beta k^2)%$ and $
\ \text{and} \ U_K(\bm{k})=2\alpha/(\alpha^2-\beta^2k^2)$$
respectively. For symbolic convenience, from now on the vector argument $\bm{k}$ will be replaced with its module $k$ by isotropicity consideration. % \ \text{respectively.}
$U_K(k)$ for 3D, for instance, can be derived immediately from the Gibbs distribution $$\sim \exp\{-(\alpha \mathcal{E}_K+\beta \mathcal{H}_K)\},$$
where $\alpha$ is the temperature parameter associated with energy and $\beta$ with helicity (enstrophy in 2D).\footnote{We adopt K73 notations and definitions: %spectral densities are defined that
%\begin{center}
$\mathcal{E}_K=\tilde{\sum_{\bm{k}}} U_K(k)
%\xrightarrow [ limit]{\text{continuous}}
\to \tilde{\int} dk 4\pi E_K(k)$ and $
%\ \text{and} \
\mathcal{H}_K=\tilde{ \sum_{\bm{k}} } Q_K(k)
%\xrightarrow [ limit]{\text{continuous}}
\to \tilde{\int} dk 4\pi H_K(k)$
%\end{center}
in the continuous-$k$ limit, where $\tilde{\bullet}$ implies restricting to the subset of surviving modes and $E(k)=k^{2} U_K(k)$ and $H(k)=k^{2} Q_K(k)$ are the 1D spectra. We will always use $\alpha$ for energy related temperature parameter and $\beta$ for helicity. Self-evident indexes, such as $M$ for ``magnetic'', when necessary for discrimination, will be added to $\beta$, $U$, $Q$, $\mathcal{E}$ and $\mathcal{H}$ etc. And, for simplicity we will always use Gibbs ensemble calculation and will not repeatedly formulate and explain it.
The general results of this paper are not affected by the differences between an infinite domain and a finite cyclic box, so we may switch between these two descriptions, depending on which one is more convenient. Difference of a factor of $2$ may arise, depending on how one treats the realizability condition (see below) and the invariant(s) (summation over the whole or half of the wavenumber domain etc.), and yet another freedom about the sign of helicity is of one's free choice. Also, spectra in this paper may be obtained in other approaches, such as %that with
finding the stationary solution of the master equation with the properties of vertices relevant to the conservation laws, which may avoid explicitly resorting to the Gibbs distribution (Private communication with E. A. Kuznetsov).}
K67 showed that low enstrophy state in 2D corresponds to a negative $\alpha$, indicating condensation of energy at smallest $k$ with a roughly (smoothed) %``L'' %
{\huge $\llcorner$}
shape spectral density: {\it c.f.}, similar MHD figures in \citet{FrischETC75}. In 3D, there is no such negative temperature state ($\alpha >0$ due to the \textit{realizability condition} from the positive definiteness of the quadratic form $\alpha \mathcal{E}_K+\beta \mathcal{H}_K$) but only {\huge $\lrcorner$} shape spectral density, and low helicity state corresponds to vanishing $\beta$ and equipartition of energy. By statistical consideration of the tendency of the interacting modes to relax towards the equilibrium state, inverse energy cascade was then argued for 2D but disputed for 3D. Note also that, as argued by \citet{LPPprl02} for 2D turbulence, in some situations turbulence is actually not far from absolute equilibrium. One may even imagine a two-step scenario that the system first reached equilibrium which was then broken with the quantity, say, energy, being removed from the scales of concentration. Such a thought experiment makes good sense when the thermalization eddy turn over time scale is reasonably %but not extremely
smaller than that of the equilibrium-breaking mechanism, such as that due to (hypo)viscosity.
In modern terminology, we may say that the system is driven by the \textit{entropic force} towards the maximum entropy state.
%\subsection{Chiroid}

\subsection{Helical (wave/mode) representation}\label{sec:HR}
For a 3D transverse vector field, such as the velocity $\bm{u}$, vorticity $\bm{\omega}=\nabla\times\bm{u}$ and the transverse component of vector potential $\bm{A}$ of magnetic field $\bm{B}$\footnote{Note that we have used Coulomb gauge $\hat{\bm{A}}\cdot \bm{k}=0$, so $\hat{i}\bm{k}\times[\hat{i}\bm{k}\times\hat{\bm{A}}(\bm{k})]=k^2\hat{\bm{A}}(\bm{k})=\hat{i}\bm{k}\times\hat{\bm{B}}(\bm{k})$: The longitudinal component of $\bm{A}$ with whatever gauge is not involved in the relevant calculations, so Coulomb gauge is not really necessary but brings symbolic convenience.
Note also that helical \textit{wave}, with the time argument included, is \textit{truly} chiral in the sense of Barron \citep[see, {\it e.g.},][]{CintasViedma12}.}
etc., in a cyclic box of volume, say, $\mathcal{V}=(2\pi)^3$, the helical mode/wave representation in Fourier space reads \citep{Moses71}%\citep[see, {\it e.g.,}][]{W92}
\begin{equation}\label{eq:FourierHelical}
\bm{v}=\sum_{\bm{k}} \hat{\bm{v}}(\bm{k})e^{\hat{i}\bm{k}\cdot \bm{r}}=\sum_c \bm{v}^c=\sum_{\bm{k},c} \hat{\bm{v}}^c(\bm{k}) e^{\hat{i}\bm{k}\cdot \bm{r}}=\sum_{\bm{k},c} \hat{v}^c(\bm{k})\hat{\bm{h}}_c(\bm{k})e^{\hat{i}\bm{k}\cdot \bm{r}}.
\end{equation}
%That is, at each $\bm{k}$ we use the same helical mode bases (this is important when cross correlation is involved but might be relaxed for some particular cases) and we have used Coulomb gauge for $\bm{A}$.
Here $\hat{i}^2=-1$ and $c^2=1$ for the chirality indexes $c=$ ``+'' or ``-''. The helical mode bases (complex eigenvectors of the curl operator) have the following properties $\hat{i}\bm{k}\times \hat{\bm{h}}_c(\bm{k})=ck\hat{\bm{h}}_c(\bm{k})$, $\hat{\bm{h}}_c(\bm{-k})=\hat{\bm{h}}_c^*(\bm{k})=\hat{\bm{h}}_{-c}(\bm{k})$ and $\hat{\bm{h}}_{c_1}(\bm{k})\cdot\hat{\bm{h}}_{c_2}^*(\bm{k})=\delta_{c_1,c_2}$, the Euclidean norm. %$\hat{\bm{h}}_{c_1}(\bm{k}_1)\cdot \hat{\bm{h}}_{c_2}^*(\bm{k}_2)=0$ if $c_1\ne c_2$ or $\bm{k}_1\ne \bm{k}_2$.
A relation used for numerical computation, such as the numerical experiments with the (pseudo-)spectral method using the various truncation schemes to be discussed in \S\ref{sec:HD}, is $\hat{\bm{v}}^c(\bm{k})%\triangleq \hat{v}^c(\bm{k})\hat{\bm{h}}_c(\bm{k})
=\hat{\bm{v}}(\bm{k})+c\hat{i}\bm{k}\times \hat{\bm{v}}(\bm{k})/k$ \citep[see, {\it e.g.},][]{Lesieur90book,MelanderHussainPoF93}.

Here the new element in the theoretical formulation of the absolute equilibrium problem lies in viewing the system a gas of %, instead of conventional Fourier modes,
pure helical modes $\hat{v}^c(\bm{k})$, representing $\hat{v}^c(\bm{k})\hat{\bm{h}}_c(\bm{k})e^{\hat{i}\bm{k}\cdot \bm{r}}+c.c.$ for simplicity, {\it i.e.}, the chiroids {\it a la} \citet[][]{Kelvin1904}, as the working `momentoids'.
Corresponding densities can be defined accordingly: For instance, \textit{mean} magnetic energy $\mathcal{E}_M=\tilde{\sum}_{c,\bm{k}}U_M^c(k)$ and helicity $\mathcal{H}_M=\tilde{\sum}_{\bm{k},c}Q_M^c(k)$ with
\begin{equation}\label{eq:QU}
U_M^c(k)=ckQ_M^c(k)=\langle|\hat{B}^c(\bm{k})|^2\rangle/2,
\end{equation}
with a reversed factor of $k$ for the kinetic case due to the difference between magnetic and kinetic helicities by definition,
where $\langle \bullet \rangle$ denotes the mean, per unit volume or in the statistical sense.
Inserting Eq. (\ref{eq:FourierHelical}) back into the rugged invariants, $\mathcal{E}$ and $\mathcal{H}$, constraining the statistical ensemble, the Gibbsian one used for our calculations, we can obtain the chirally split densities $U^c(k)$ and $Q^c(k)$ which present finer physical structures than the mixed ones
\begin{equation}\label{eq:QUc}
U(k)=U^+(k)+U^-(k) \ \text{and} \ Q(k)=Q^+(k)+Q^-(k):
\end{equation}
We remark that this result should be perceived in two perspectives. One is that the AE spectra of pure helical modes of each chiral sector present separately, independent of the existence of the other one, {\it i.e.}, whether the other sector is truncated or not for some $\bm{k}$s, since the truncation can be performed arbitrarily on the chiroids, except that Hermitian symmetry should be kept like the classical Fourier truncation; the other perspective is that the spectra are chirally decomposed into two sectors, if both exist, {\it i.e.}, the truncations for both sectors are symmetric.
Note that $|Q_M(\bm{k})|\le U_M(\bm{k})/k$, so the purely helical mode is called \textit{maximally helical} \citep[see, {\it e.g.},][]{k73}.
Other derivatives such as the relative helicity %defined by the total helicity and energy densities
$|kQ_M(k)|/[U_M(k)]=k\sum_c Q_M^c(k)/\sum_c U_M^c(k)$ \citep[][with a reversed factor of $k$ as in Eq. \ref{eq:QU}]{k73} can be used for quantifying the degree of chirality.
Only when the modes have the same wavelength and are \textit{homochiral}, i.e., all with same handedness, can we see the physical-space field resulting from their superposition is Beltrami, {\it i.e.}, $\nabla\times\bm{v}=\gamma \bm{v}$ with constant $\gamma$, {\it force free} for magnetic field, in which case, nonlinearity is typically depleted completely.

\subsection{Statistical ensembles of truncated chiroids}\label{sec:BoTCS}
When the system is reduced to the dynamics of the helical modes, one can then consider various truncations directly on such chiroids \citep[][]{W92,bmt12}.
Detailed triadic interactions of the helical Fourier modes %chiroids
in various hydrodynamic-type models, such as HD, %Shell model,
MHD, EMHD and Hall MHD etc., have been closely looked into by different authors \citep[{\it e.g.},][]{W92,LessinnesPlunianCarati,GaltierBhattacharjee03,GaltierHMHD06}. %have respectively looked into the d % respectively. %Noteworthily,
And, relevant details of the AE calculation have been well described in the literature %(especially the series of papers by Kraichnan on HD and by Frisch et al.'s on MHD)
and do not require any further elaboration here. For instance, it is routinary %to write down the chirally decomposed dynamical equations, especially the triadic interaction geometrical coefficients, and
to check Liouville theorem and rugged conservation properties after Galerkin truncation of the helical modes, which is true for all the models studied here \citep[for HD, {\it c.f.}, equations 7 and/or 9 of][]{W92}. To be a bit more definite but without loss of generality, using indexed $y$s for the variables related to the real and imaginary parts of the active chiroids $\hat{v}^c(\bm{k})$, we can write down the dynamical equations as
\begin{equation}\label{Y}
    \partial_t y_n=\sum_{l,m}Y_{lmn}y_ly_m,
\end{equation}
with $Y_{lmn}$ satisfying some specific symmetries to assure the detailed conservation of energy and relevant helicity(ies)  and Liouville theorem. Note that for cases with linear terms of the original variables on the right-hand side, such as the 3D gyrokinetics \citep{gkaeOLD}, a simple linear transformation of variables reduces them to this form which is formally the same as the well-known classical Fourier Galerkin truncated Euler case. Readers can go to the Appendixes for more discussions on the \textit{detailed conservation laws, dynamical and topological aspects, and, tacit assumptions about ergodicity or mixing etc. in the statistical considerations}.

\subsection{Plan}
We progressively perform a minimal but systematic investigation, with different emphases, of EMHD with formally pure magnetic field dynamics in \S\ref{sec:EMHD}, HD in \S\ref{sec:HD}, and, single-fluid (\S\ref{sec:sfMHD}) and two-fluid (\S\ref{sec:tfMHD}) MHDs. \S\ref{sec:EMHD} discusses mainly the natural chiral selection for inverse magnetic helicity transfer; \S\ref{sec:HD} concentrates on the chirally asymmetric truncation effects, \S\ref{sec:sfMHD} on new insights to the classical dynamo issue, \S\ref{sec:tfMHD} on the two-fluid effects. Although many of the discussions in the (sub)subsections, such as the asymmetric truncations of the two chiral sectors for some $\bm{k}$s in \S\ref{sec:HD}, can be carried over to other (sub)subsections, {\it mutatis mutandis}, to get some relevant new insights, we won't detail such obvious points.
The general purpose is to lay out the basic AE %with traditional Galerkin truncation
as a first step to explore some fundamentals of turbulent transfers, especially for a comprehensive basic understanding of the relevant helicity effects. %\footnote{
It should be pointed out that there have been many other interesting AE-relevant investigations for different dynamical models and on various specific physical issues, the important one of which is relevant to a space uniform magnetic field and anisotropic fluctuations and has been continuously attacked %by Montgomery, Matthaeus and collaborators
in the last several decades. Extra particular studies should be done, though $k=0$ mode can formally be included in the calculations and brief relevant remarks will be offered at the appropriate circumstances. %}
The focus is the most basic new insights attached to the chirally decomposed AE, with which we will revisit the most relevant studies, rather than any other specific turbulence (closure) theories, such as the wave turbulence theories studied by Galtier and collaborators, though our results may be used as benchmarks of relevant analytical or numerical treatments.

In summary, an incompressible hydrodynamic-type system can be reduced to the dynamics of pure helical modes, the ``chiroids'' {\it a la} Kelvin, with helical wave/mode representation. Left- and right-handed sectors of the absolute statistical equilibrium spectra are split. Either sector may present without the necessity of the existence of the other, {\it i.e.}, unichirally with asymmetric Galerkin truncation. One sector can dominate around its positive pole(s) with corresponding net helicity, providing a unique chirality selection and amplification mechanism. Chirally truncated systems preserve Moffatt's topological interpretation of helicity due to the detailed incompressibility for each chiroid. We obtain new insights about chirality selection and amplification, and, spectral transfers of turbulence of various neutral-fluid and plasma systems. For instance, one-chiral-sector-dominated states are naturally supported by magnetohydrodynamics (MHD) absolute equilibria with magnetic helicity, and homochiral Euler system allows negative temperature states which were excluded by Kraichnan for the non-homochiral case. A major purpose is to make a systematic comparison of the effects of various helicities, for finding genericities and specificities, and we clarify the special role of magnetic helicity for turbulent inverse magnetic helicity transfer/cascade by analyzing the electron MHD, with only magnetic field, and the two-fluid MHD, with the combination of various helicities in a symmetric way.

As said, it is not necessary to elaborate the calculations repeatedly, thus the following presentation will mainly consist in a set of brief backgrounds and theoretical frameworks, discussions of our results with careful comments on relevant studies and new explanations of documented data. Readers are suggested to go directly to the (sub)subsection(s) for the interested model(s)/topics first and then, before trying to read the analyses of others, to \S\ref{sec:Conclusion} for further discussions, where not only the major results are summarized, but also the genericities and differences are extracted by comparisons across different models.

\section{One-chiral-sector-dominated absolute equilibrium and turbulence states}

Basic (rules of) notations, definitions and calculation techniques follow \S\ref{sec:hae}. Readers are assumed to be familiar with the relevant ideas and techniques of K67 and K73 summarized there. We do not repeat the further detailed simple calculations of the spectra following K73 and \citet{FrischETC75} which will actually be found to be greatly simplified, since much, for single- and two-fluid MHDs, or all, for EMHD and HD, of the diagonalization work and the solenoidal constraints, have already been performed from the beginning with the helical representation while constructing our ensemble.

\subsection{Pure magnetic fluid}\label{sec:EMHD}
EMHD equation $$\partial_t\bm{B}+\nabla\times\big{[}(\nabla\times\bm{B})\times\bm{B}\big{]}=0$$ formally involves only the magnetic field and may be called pure magneto-dynamics.
This fluid model %also corresponds to the case where Hall effect dominates in plasma, in particular in Hall MHD \cite{Turner86}, and
corresponds to the small electron skin depth $d_e\ll 1$ limit of the more general case, which we will discuss later, and is relevant to helicons or whistler waves in solid conductor, including neutron star's solid crust, atmosphere etc \citep[see, e.g.,][and references therein]{BiskampETC99,GaltierBhattacharjee03}. Note that $\bm{B}$ is ``frozen in'', by definition, to the electron fluid velocity $\bm{u_e}=-\nabla\times\bm{B}$.
Rugged invariants are magnetic energy and helicity:
$$\mathcal{E}_M=\frac{1}{2\mathcal{V}}\int \bm{B}^2 d^3\bm{r}=\frac{1}{2}\sum\limits_{\bm{k},c}|\hat{B}^c(\bm{k})|^2 \ \text{and} \ \mathcal{H}_M=\frac{1}{2\mathcal{V}}\int \bm{A}\cdot \bm{B}d^3\bm{r}=\frac{1}{2}\sum\limits_{\bm{k},c}c|\hat{B}^c(\bm{k})|^2/k.$$
The two chiral sectors of the AE spectral densities ({\it c.f.}, \S\ref{sec:hae}) of energy and helicity are then
\begin{eqnarray}
 U_M^c(k) %&& =\frac{1}{\alpha+\frac{c}{k}\beta}
 =k/(c \beta+\alpha k)%, %\label{eq:EmhdUB}\\
\ \text{and} \   Q_M^c(k) %&&
%= cU_M^c(k)/k
=1/(\beta+c\alpha k)= cU_M^c(k)/k \label{eq:EmhdQM}.
\end{eqnarray}

From the above spectral relations, just as K67, but with energy playing the role of enstrophy there, a low energy state corresponds to condensation of $Q$ at smallest wavenumbers, close to the positive pole $k_p=-c\beta/\alpha$ of one of the chiral sectors, say, $c=+$, with $\beta<0<\alpha$. The implication for turbulence is inverse helicity and forward energy transfers%, when scale separation is satisfied
.
In principle, as long as there is net helicity with $\beta\ne 0$, one chiral sector can dominate at large scales, {\it i.e.}, OCSDS around the positive pole, though commonly done in experiment to provoke EMHD turbulence is to impose a background magnetic field \citep[see, e.g.,][]{StenzelJGR99} which breaks the skin effect and guides the waves.
If helicity is injected at some intermediate $k$, we should see dominant inverse helicity and forward energy transfers.
%, which, i
If the transfers in these two regimes are approximable by self-similar local cascades and that suitable for simple dimensional analysis, energy spectra follow $k^{-5/3}$ and $k^{-7/3}$ scalings: \citet{BiskampETC99} first proposed and presented in slightly different situations such scalings, and their Fig. 8b with electron skin depth $d_e\ll 1$ does correspond to the forward energy cascade of our case.
%We emphasize that, %like (but opposite to) the isolated pure helical wave system of 3D Navier-Stokes of Biferale et al. \cite{bmt12},
The scale separation between the dominant dynamics of the two cascade quantities of EMHD is weaker than 2D fluid turbulence, in the sense that the spectral ratio of each chiral sector is at most $k$, instead of $k^2$ of the latter. So, at finite Reynolds numbers, cascade of either definitely is accompanied by stronger (than 2D turbulence) leaking of the other.
%Like pure hydrodynamics (HD), i
The subdominant energy transfer, accompanying the inverse helicity transfer and vanishing at high Reynolds number limit, should not be recognized to be the genuine inertial cascade.
%just as in single-fluid MHD (\S\ref{sec:sfMHD}) where magnetic energy is not even conserved.
%\footnote{
One should be particularly careful for the decaying case which is exactly the nice simulation by \cite{Cho11},
who, by ``inverse energy cascade'', meant merely the backward shift of the peak of his energy spectrum, which is not genuine in connect to the conventional notion of inertial cascade and which is not in conflict with our statement of forward energy cascade (Private communication). %}

Our result indicates a largest-scale nearly force-free magnetic fields.
%: Only when the modes have the same wavelength and are homochiral (with the same handedness), can we see the physical-space field resulting from their superposition is Beltrami ({\it force free} for magnetic field).
For the discrete-$k$ case, the smallest-$k$ modes contain most of the energy, so the whole global structure may appear to be roughly Beltrami, with smaller-scale ``turbulent'' fluctuations.
%Note that in the 1950s, force-free field was intensively studied with several variational formulations, some of which are purely magnetic \citep[][and references therein]{Woltjer59}.
Note that completely force-free fields, instead of ours with the scale-dependent degree of chirality measurable by the relative helicity \citep{k73}, were obtained with several variational formulations in 1950s: We will come back to this in \S\ref{sec:sfMHD} for single-fluid MHD.

The basic feature, concerning the issue of magnetic helicity inverse transfers/cascades, of the EMHD results in the above is also central to other MHD models with magnetic field.
Some brief remarks for the finite-$d_e$ (which is used for scale normalization here) general EMHD model, $$\partial_t(\nabla^2\bm{B}-\bm{B})+\nabla\times\big{[}(\nabla\times\bm{B})\times(\nabla^2\bm{B}-\bm{B})\big{]}=0,$$ are in order. The ``frozen-in'' generalized vorticity is
\begin{equation*}
\nabla\times \bm{P}_e %(\bm{u}_e-\bm{A})
=\nabla\times \bm{u}_e-\bm{B} \ \text{with} \ \bm{P}_e=\bm{u}_e-\bm{A}.
\end{equation*}
The rugged invariants are now total energy and \textit{generalized} helicity\citep{BiskampETC99} $$\mathcal{E}=\frac{1}{2\mathcal{V}}\int (\bm{B}^2+\bm{u}_e^2) d^3\bm{r}, \ \mathcal{H}_G=\frac{1}{2\mathcal{V}}\int \nabla\times \bm{P}_e \cdot \bm{P}_e d^3\bm{r},$$ resulting in $$U_M^c(k) =k/\{(k^2+1)[c\beta (k^2+1)+\alpha k]\},$$ which complicates the quantitative transitional spectral behaviors ({\it c.f.}, \S\ref{sec:tfMHD} for more general discussions for similar situations in two-fluid MHD). In the other limit regime of scales much smaller than $d_e$, or in another word, when $k\gg1$ and that $k^2+1$ can be replaced by $k^2$, $k^2U_M^c(k) \to 1/(c\beta k+\alpha)$, the same as the pure HD case, supporting both energy and helicity cascading forwardly
\citep[{\it c.f.}, Fig. 8a of][%and see \S\ref{sec:HD} for detailed discussions
]{BiskampETC99};
\footnote{Note that, for convenience, in this regime one may want to study magnetic enstrophy $W$ defined through $W(k)=k^2U_M^c(k)$ and the other quantity $S$, which one might want to call magnetic \textit{helistrophy}, defined through $S(k)=k^2Q_M^c(k)$. Such an attempt however is conceptually not very appropriate, since neither of them are conserved quantities.}
%Practically, such as in plotting the spectra $W(k)$ and $S(k)$, it is fine, and especially in our current comparison with the HD scenario, it does be very helpful.
Magnetic helicity concentrating at or transferring to ``large'' (of course in the sense of subsidiary limit) scale can only be obtained with imposed asymmetric truncation between the two chiral sectors, such as that leaving only the $c=+$ sector \citep[][and see \S\ref{sec:HD} for the discussions]{W92}. This should not be surprising, since electron kinetic fluid flow dominates in this limit.

\subsection{Pure neutral fluids}\label{sec:HD}
For the classical incompressible HD, {\it i.e.}, $$\partial_t\bm{u}+\bm{u}\cdot\nabla\bm{u}=-\nabla p,$$ where the pressure $p$ can be eliminated by $\nabla\cdot\bm{u}=0$,
the rugged invariants are kinetic energy and helicity $$\mathcal{E}_K=\frac{1}{2\mathcal{V}}\int \bm{u}^2 d^3\bm{r}%=\frac{1}{2}\sum\limits_{\bm{k},c}|\hat{u}^c(\bm{k})|^2
, \ \mathcal{H}_K=\frac{1}{2\mathcal{V}}\int \nabla\times\bm{u}\cdot \bm{u}d^3\bm{r}%=\frac{1}{2}\sum\limits_{\bm{k},c}ck|\hat{u}^c(\bm{k})|^2
,$$
which lead to %the AE results:%
the densities of separate chiral sectors:
\begin{eqnarray}
 U_K^c(k) %&&
 =1/(\alpha+c\beta k), \ %\label{eq:UK}\\
   Q_K^c(k) %&&
   %=\frac{c k}{\alpha+c \beta k}
   = c k U_K^c(k). \label{eq:QK}
\end{eqnarray}
Note that the above spectra can not be considered to be simply the decomposition (into two chiral sectors) of K73 densities ({\it c.f.}, \S\ref{sec:hae}) which are not valid when there is asymmetric truncation between the two chiral sectors of some $\bm{k}$, that is, when only one of the chiral sectors of some $\bm{k}$ is truncated to be unichiral. For example, if there is no cancelation at some $k$, one can not derive from Eq. (\ref{eq:QK})
\begin{eqnarray}
\alpha=[U_K^-(k)+U_K^+(k)]/[2U_K^+(k)U_K^-(k)], \ \beta=[U_K^-(k)-U_K^+(k)]/[2kU_K^+(k)U_K^-(k)]
\label{eq:abK}
\end{eqnarray}
which are in particular not true for any $k$ in the homochiral system with only one, say, the positive chiral sector. In such homochiral case with $c=+$, $\alpha >0$ is not required by realizability condition (\S\ref{sec:HD}) and Eq. (\ref{eq:QK}) shows that the low helicity state corresponds to a negative $\alpha$ with a sharp peak at the lowest modes $k_{min}>k_p=|\alpha/\beta|$ close to the positive simple pole $k_p$. Such {\huge $\llcorner$}-shape energy spectral density, just as K67, indicates inverse energy and forward helicity dual transfers \citep[][who also aruged for this with his ``instability assumption'']{W92} with Kraichnan's argument of the tendency of relaxation towards equilibrium, as numerically realized by \citet{bmt12} with remarkable quality. Note however that such HD scenario does not work in vanishing-$d_e$ EMHD in \S\ref{sec:EMHD}, neither for other more complicated MHD models as will be studied in the next section, which, for instance, when truncated to be homochiral, presents no drastic change of transfer/cascade direction; this is because the large-scale magnetic helicity concentration of OCSDS with symmetric truncation and that of the homochiral state coincide.

\subsubsection{OCSCSs in the sense of fluxes: ``Second order'' OCSDSs}
If both sectors present for every $\bm{k}$, %\footnote{In reverse to truncation/decimation, we may call such a procedure ``medication''.}
\textit{only} {\huge $\lrcorner$} shape spectral density dominated by one of the sector around the peak is allowed by the realizability conditions $\alpha>0$ and $k_{max}<\alpha/|\beta|%=k_p
$ from the positive definiteness of the quadratic form: {\it c.f.}, \S\ref{sec:hae}. %K73 thus excluded the possibility of inverse cascade ``driven'' by natural relaxation.
Large-scale condensation mechanism is absent, thus inverse cascade in HD generally needs other special treatments as reviewed by \citet[][]{YangWu10} and \citet{bmt12}. %Note that, such large-$k$ peak is dominated by one of the chiral sector with its positive simple pole.
Although, unlike at large scales, normal dissipation would devastatingly ruin such small-scale explicit OCSDS AE structure, some residuals of such intrinsic nonlinearity effects may persist. Indeed, the fluxes of the two chiral sectors reported by \citet{CCE03} do show systematic differences: According to the working conditions, their Figs. 2--5 correspond to the case with positive helicity, that is, the positive pole belongs to the positive sector with negative $\beta$. Dominance of the positive sector of AE spectrum indicates that nonlinearity should support the transfer of this sector to be more persistent, consistent with the results and analyses of \citet{CCE03}.
We may also call them ``implicit'' OCSDSs, or ``second order'' OCSDSs, in the sense of dominance of energy and helicity fluxes of one sector, as signatured by the lower panels of their Figs. 4 and 5.
Such second-order OCSDSs may be viewed as yet another special evidence, besides the isotropization \citep{Lee1952} and bottleneck \citep{FrischETCprl08}, of the \textit{persistence of thermalization} around the end of inertial range.
The large-$k$ viscous effect efficiently restores the reflexion symmetry, %of restoration of reflexion symmetry, {\it i.e.}, equipartition between the two sectors% is very strong,
and the degree of local-in-$k$ chirality measured by relative helicity vanishes as $k^{-1}$ throughout the inertial range with accurate $k^{-5/3}$ scaling exponents for both energy and helicity \citep{k73}.
The large-$k$ pole effect of one chiral sector nevertheless provides a prototype for other similar possible physics ({\it c.f.}, \S\ref{sec:tfMHD}) in more complex situations, furthermore it might be possible to find its stronger activity in a non-Newtonian fluid such as the (dilute) solution of chiral polymers, say, deoxyribonucleic acid (DNA), in a normal fluid where some kind of resonance between one chiral sector of the fluid motion and the polymer's chiral structure/activity could happen; see more relevant discussions in \S\ref{sec:GR}.
%Note that in the EMHD case, the realizability condition is $\alpha>0$ and $k_{min}>|\beta|/\alpha$, and the {\huge $\llcorner$} shape spectrum is the only possibility.

\subsubsection{OCSDSs with special truncation schemes and ``smooth'' transitions of energy transfer directions% and discussion of rotating turbulence
}\label{sec:forwardinverse}
Negative temperature state emerges with dramatic physical indications for the homochiral case. Now, if we add just one pair of conjugate modes, with opposite helicity, negative temperature is then excluded, by the nonnegativity of $U^c_K$. Naturally, a kind of ``phase transition'' %\footnote{A traditional rigorous definition of \textit{phase transition} is not trivial here, since we add the mode in a discrete way, which however is not important for our discussion here and below: We merely use the terminology, with quotation marks, in a loose way, to highlight the feature of sharp change.}
is happening in the temperature parameters, since the negative temperature must jump to some positive value. Then with superficial impression from the above analysis and from K73 one would tend to expect similarly a sharp transition of inverse energy cascade to forward energy cascade. But, such a superpowerful potential of a single alien chiroid would be shocking. Thus we need to look into the corresponding absolute equilibrium states by starting with a clean pool of homochiral $c=+$ modes and put aliens into it. \textit{It turns out that there should be no phase-transition-like behavior concerning cascades and energy can ``smoothly'' switch from completely-inverse to partly-inverse-and-partly-forward and to completely-forward cascades, depending on how (many) aliens are put into the pool. For the general truncations of chiroids with asymmetry but not homochirally, K73's argument for excluding inverse energy cascade still does not simply work, and large-scale concentration of energy, indicating inverse energy transfer in turbulence, can exist without a negative temperature state.}
To see this, suppose we have only one alien chiroid, {\it i.e.}, one negative-helicity mode, at the objective \textit{condensation wavenumber}, $k_c=k_{min}$ much smaller than the injection wavenumber $k_{in}$, and that negative temperature state is excluded for implying a conventional inverse energy transfer/cascade argument {\it a la} K67. %\footnote{If isotropicity is pursued, we need to fill the whole shell $|\bm{k}|=k_{min}$ with aliens, but the number of modes is not essential to our statistical discussions here (though possibly relevant to realization of dynamical transfers) and is omitted for simplicity of notation and wording.}
Would energy abruptly turn to cascade forwardly, or there should be a transitional behavior? To get illumination, one may apply to the arguments and thought experiment presented at the end of \S\ref{sec:HTaAE}.
Consider Eqs. (\ref{eq:QK}) and (\ref{eq:abK}) with $\alpha>0$ and $\beta>0$, and suppose energy is injected at some intermediate $k_{in}$.
This alien mode may help transfer extra positive helicity nonlocally to small scales, by which, though, its own amplitudes of (negative) helicity and energy would have to increase; this is because for the excitation of any mode in $k>k_{in}$ the injected helicity at $k_{in}$ is not enough to support it, just by the relation $Q_K(k)=kU_K(k)$ for pure helical mode, thus some extra positive helicity should be provided from $k<k_{in}$, which is facilitated by the excitation of negative helicity in the alien mode. The growth of this alien mode is allowed, even in the sense of complete AE, with $\alpha/\beta$ approaching $k_c^+$, {\it i.e.}, from above:
\begin{equation}\label{eq:nonk73}
\alpha/\beta \to k^+_c  \ \text{and that} \ U_K^-(k_c)\to \infty, \ \text{turning $k_c$ into the pole $k_p$}.
\end{equation}
%justifying our naming $k_c$ the \textit{condensation wavenumber}.
Note that $k$ can be larger than $\alpha/\beta$ even though $U_K^-(k)=1/(\alpha-\beta k)$, since the alien mode in this sector is restricted below
%\sout{\xout{
$k_c$
%}}
%\textbf{$\alpha/\beta$}
, actually, as said, the $c=-$ sector of Eq. (\ref{eq:QK}) being only for $k_c=k_{min}$ now, unlike the traditional symmetrically truncated system with $\alpha/\beta>k_{max}$ as discussed by \citet{k73}; or, in another word, \citet{k73} could not let $\alpha/\beta$ approach $k_c$ from above, because he had other larger-$k$ alien modes which otherwise would have negative energy for $k>\alpha/\beta$. \textit{Thus, adding such a single alien excludes the negative-temperature state, but the single alien can carry the energy condensating there as a particular form of OCSDS.}%\footnote{Note that the physical field carried by this alien, by the definition of pure helical mode \citep[{\it c.f.}, Eqs. 3 and 4 of][]{k73}, would be ``quasi-two-dimensional'' without dependence on the coordinate, say, $\bm{z}$, along which both the velocity and vorticity are aligned, perpendicular to $\bm{k}_c$ and that would present columnar structures.}
%\sout{
The $k$-distribution of energy of such a state does not differ from that of the homochiral negative temperature state too much.
%}
And, one probably can carefully design a simulation by adding an alien to the smallest $k$ of \citet{bmt12}'s simulation and still get inverse transfer of energy.
When we gradually increase the number of aliens starting from $k_{min}$ to the condensation wavenumber $k_c\ge k_{min}$, there may or may not be a nonzero forward nonlocal transfer of energy to small scales in the infinite Reynolds number limit; or, in another word, depending on the strength of the nonlocal kicking of the alien modes, the energy accompanying the forward helicity transfer may or may not vanish as the wavenumber goes to infinity: It may not be impossible that, on average, a single alien could remotely ``kick'' the small scales by ``stirring'' up vortex stretching over the field, making the solution rougher and that causing some finite energy dissipation.
%[An inertial range with constant helicity flux of the conventional cascade scenario is neither assured with such a special non-local interaction mechanism, and the relation we have between the energy and helicity \textit{spectral transfer functions} of each chiral sector $T^c_E(k)=ckT^c_H(k)$ is not enough to determine it.]
%If there is, it is just that the injected energy is partitioned to be transferred to two opposite directions.
With more aliens, forward helicity transfer will be more. The injected energy must gradually be partitioned to be transferred to two opposite directions simultaneously in the infinite Reynolds number limit. And, to transit to a state with all energy completely cascading forwardly, sufficient aliens must be added to the regime of $k$ larger than $k_{in}$, since if all aliens are added only to the regime below $k_{in}$, the nonlocal interaction across $k_{in}$ for transferring helicity to larger $k$ must be accompanied with some transfer of energy to $k<k_{in}$, as is also indicated by analyzing the corresponding absolute equilibria [{\it c.f.}, statement (\ref{eq:nonk73}) for $k_c<k_{in}$]. Regarding the cascades and/or nonlocal transfers in the infinite Reynolds number limit, the approach we have taken is \textit{tuning the degree of the regularity of the solutions to be appropriately ``dissipative'' by manipulating the population of the helical modes}. To our best knowledge, for the full 3D Navier-Stokes, actually even for 2D, there is not yet satisfying mathematical theory for the cascade statements in the infinite Reynolds number limit. But, it appears that we may use the absolute equilibrium states to obtain at least some fine and clear physical-picture intuitions for such an approach.
%For the pure magnetic dynamics of EMHD in the last subsection, we can also do similar asymmetric truncation, by limiting the chiral sector with positive pole only to large $k$ so that $k_p$ can be approached by $|\beta/\alpha|$, but from below, which we have not yet found particularly interesting point to elaborate further.

\subsubsection{HD summary}
We have come from \textit{magneto-dynamics} in \S\ref{sec:EMHD} to this \textit{hydro-dynamics} which, with the great substantiation for transition to \textit{magneto-hydro-dynamics}, needs a summary:
\begin{enumerate}
  \item \citet{k73}'s argument for forward energy and helicity cascades can be refined to find the relevance of his absolute equilibrium with the energy/helicity fluxes reported by \citet{CCE03}, as a kind of ``second order'' OCSDS;
  \item as shown by our chirally split spectra, his argument however is not for the homochiral truncation where the negative temperature state, indicating inverse energy transfer/cascade, is allowed;
  \item such a ``sharp'' change of cascade scenario is not shared by the pure \textit{magneto-dynamics};
  \item The cascade transition is not really ``sharp'' but has a ``smooth'' dependence on how alien modes are added to the homochiral pool.%\textbf{, which is relevant to rotating turbulence}.
\end{enumerate}

\subsection{Magnetised fluids}
In principle, there can be many fluid models for describing different subsets of the kinetic phase space of plasma dynamics \citep[see, {\it e.g.}, a very limited list in a review by][]{SchekochihinETC09}. Here we study the classical single-fluid and the most general two-fluid MHDs. From the plasma physics point of view, two-fluid MHD is for a more complete description of the kinetic effects for the dynamics/scales between those of EMHD and single-fluid MHD. Two-fluid MHD presents various helicities in a unified way.

\subsubsection{Single-fluid MHD}\label{sec:sfMHD}
As introduced in \S\ref{sec:hae}, \citet{MeneguzziPRL81} found with direct numerical simulations that ``the large-scale $\bm{B}$ is mostly force free and produces only very little large-scale motion,'' with the relative magnetic helicity density $|kQ_M(k)/U_M(k)|$ being close to $1$, nearly maximally helical%, as is how we recognize the OCSDSs in \citet{PouquetETC76}
; and, recently, \citet{BrandenburgDoblerSubramanian02} explicitly pointed out, by their Fig. 21 with postprocessing using helical decomposition, that the simulation with similar setup also present such OCSDSs. We now turn to explain such findings with the classical single fluid MHD equations
\begin{eqnarray*}
 \partial_t \bm{u}&=& -(\bm{u}\cdot \nabla)\bm{u}+(\bm{B}\cdot\nabla)\bm{B}-\nabla (p+B^2/2), \\% \label{eq:MHDu} \\
 \partial_t \bm{B} &=& -(\bm{u}\cdot \nabla)\bm{B}+(\bm{B}\cdot \nabla)\bm{u}, \ %\\%\label{eq:MHDb}  \\
\text{where} \ %&&
\nabla\cdot\bm{u} = 0 \ \text{and} \ \nabla\cdot\bm{B}=0.%\label{eq:MHDincompressibility}
\end{eqnarray*}
%\end{eqnarray}
Rugged invariants are three \citep[see, {\it e.g.},][]{Woltjer59,FrischETC75}, the energy, magnetic helicity and cross helicity $$\Big{\{}\mathcal{E}=\frac{1}{2\mathcal{V}}\int (\bm{u}^2+\bm{B}^2)d^3\bm{r}, \ \mathcal{H}_M=\frac{1}{2\mathcal{V}}\int \bm{A}\cdot \bm{B}d^3\bm{r} \ \text{and} \ \mathcal{H}_C=\frac{1}{2\mathcal{V}}\int \bm{u}\cdot \bm{B}d^3\bm{r}\Big{\}},$$
which, together with Eq. (\ref{eq:FourierHelical}), leads to ({\it c.f.}, \S\ref{sec:hae}, also for notations)
\begin{eqnarray}
U_{K}^c(k)  =\,{\frac {4(\alpha\,k+c \beta_M)}{(4\,{
\alpha}^{2}-{\beta_C}^{2})k+c 4\,\alpha\,\beta_M}}, \ %\label{eq:mhdUK}\\&&
U_M^c(k)  =\,{\frac {4(\alpha\,k)}{(4\,{
\alpha}^{2}-{\beta_C}^{2})k+c 4\,\alpha\,\beta_M}}, &&\label{eq:mhdU}\\
Q_M^c(k)  = \frac{c}{k}U_M^c(k), \ Q_K^c(k)=c kU_K^c(k), \ %\label{eq:mhdQM}\\&&
Q_C^c(k)  = \,{\frac { -2
\beta_C k}{(4\,{
\alpha}^{2}-{\beta_C}^{2})k+c 4\,\alpha\,\beta_M}}. &&\label{eq:mhdQ}
\end{eqnarray}

Similar to the statement in \S\ref{sec:HD} in comparing our spectra to those of \citet{k73}, our spectra can not be considered to be simply the decomposition of those of \citet{FrischETC75} which are not valid when there is asymmetric truncation of the two chiral sectors of some $\bm{k}$. When there is no asymmetric truncation, our spectra chirally split those of \citet{FrischETC75} following whom we start with the case of null cross helicity with $\beta_C=0$ for discussions:
$Q_M^c(k)$ with $sgn(c\beta_M)=-1$ is responsible for the condensation of $Q_M$ at small $k$, around the positive simple pole $k_p=-c \beta_M/\alpha$.
When the dynamics is dominated by the $c=+$ ($c=-$) sector, it is simply to say that large positive (negative) magnetic helicity state corresponds to a negative (positive) $\beta_M$ with a {\huge $\llcorner$} shape spectral density %, and that inverse magnetic helicity transfer,
is favored: \citet{FrischETC75} plotted such spectra, the spectral densities multiplied by $k^2$, in their Figs. 1 and 2 for illustration, by choosing several typical temperature parameters. The other sector's pole has the opposite sign and is not reachable, thus, without such a mechanism of large-scale attraction, the energy would be transferred to small scales or simply less excited. When $\beta_C$ (or $\mathcal{H}_C$) is nonzero, the prefactor before $k$ in the denominators quantitatively changes% (kinetic energy is less equipartitioned)
, but the qualitative picture is not altered. % \sout{\textbf{\textit{(probably indicating stronger inverse transfer of magnetic helicity, since the pole becomes larger)}}}.
As pointed out in \S\ref{sec:EMHD}, the large-scale nearly maximally helical state predicted by AE is close (see also next paragraph) but different to the purely force-free one intensively studied in 1950s \citep[see references in][]{Woltjer59}. The common feature with that of \citet{Woltjer59} is that the invariant cross helicity does not essentially change the large-scale nearly maximally helical physics. %\footnote{
Recently, some authors argue that cross-helicity, signature of the imbalance along and opposite to the background magnetic field, may be important to determine the (reduced) MHD forward cascade inertial range scaling exponent \citep[see, {\it e.g.},][and references therein]{PerezETC12}, which is beyond the scope of this study, although a spacially uniform $\bm{B}_0$ can be formally included in our calculation and analysis.%}

Thus, the OCSDSs in \citet{MeneguzziPRL81} and \citet{BrandenburgDoblerSubramanian02} are related to our AE spectra. Actually, \citet{PouquetETC76} had carried out systematic study of eddy-damped quasi-normal Markovian (EDQNM) MHD turbulence and nonlinear dynamo.
The data of their Figs. 4 and 5 around $k=0.16$, actually starting from the beginning of the inertial range to ever larger scales, and, of Figs. 8 and 9 around $k=0.1$% (actually also $k=2$)
, already clearly presented OCSDSs as can be seen from the values of the relative helicities computed from their figures.
Obviously, as EDQNM shares the conservation properties of the original system, it satisfies the AE spectra and that the OCSDS arguments also work.
%We thus see that, except for the explanation/prediction from a helical decomposition theory, this school were almost already here!
%The above new insight about OCSDS compared to \citet{FrischETC75} actually can be found in the numerical data documented by this group:
%Now we look back at some reported relevant simulations:
%Note that, in general, transfer with a net magnetic helicity does not necessarily mean OCSDT, as helicities of the two sectors could simply be of the same order.
We won't go too far into much more details, but just remark that the pertinent discussion of \citet[][p. 345, second paragraph]{PouquetETC76} can also be elaborated: For instance, the large-scale ensemble can be understood by AE with positive magnetic helicity, with finer chiral-sector dynamics for the different chiralities in separate scale regimes. %because when forcing or dissipation are involved the situation becomes more complicated (though our chirally decomposed AE may still be relevant to explain the forward OCSDT by the opposite sector, by simply further exploiting the magnetic helicity conservation argument.)
%There could be other (inverse-)dynamo(s) or mechanism(s) of (anti-)dynamo, but such a self-consistent simple chiral-sector statistics explanation is attractive.
We only want to remark that relaxation of magnetic helicity to the largest scales does not necessarily indicate local cascade, nonlocal transfer also is possible \citep[see, {\it e.g.}][]{BrandenburgSubramanian05}; and, there is nothing in conflict with the more mechanical reasoning, such as the alpha effect.
%But, details of transfer mechanism, such as (saturation) time scales, are in general beyond the AE calculation.
%The observed OCSDSs consistent with absolute equilibrium indicates that those large scales may actually be close to statistical equilibrium state when they are more or less saturated.

Conceptually, a reader may quickly question whether we really learn anything more from chiral decomposition than that if
helicity is large, one chirality must dominate. Isn't that obvious? The
simple pole mentioned under Eqs. (\ref{eq:mhdU}) and (\ref{eq:mhdQ}) is already present in the spectra of \citet{FrischETC75} and indeed accounts for the
accumulation of magnetic helicity at large scales. Chiral decomposition is not needed to reach this
conclusion? The answers are ``no''s.
To understand these problems, one must first understand that large net helicity does not necessarily mean OCSDS or big relative helicity.
%The reason is explicitly shown by the equalities of numbers $1001-1=1000=1001000-1000000$, where $1000$ represents a large number for the net helicity in the system not viewed with chiral decomposition, and, where $1001$ versus $-1$ on the left-hand side and $1001000$ versus $-1000000$ on the left-hand side represent respectively the positive- versus negative-sector helicities of two cases (the former of which is OCSDS and the latter of which is not, obviously.)
Now, suppose the spectra were not chirally decomposed and that the small-$k$ pole might contribute to both left- and righ-handed helicities. In such an ambiguous situation, one would not be able to conclude OCSDS as the scale approaching the pole, which was exactly what happened in the past studies mentioned in the above, sounding like just a hair's breadth though.
Interestingly, even in the extremely strong sense of ``non-helical'' state with $Q_M(k)=Q_K(k)=0$, {\it i.e.}, the two chiral sectors of both magnetic and velocity fields balancing at each wavenumber, AE seems to still support the so-called non-helical turbulent dynamo%, in the sense of \citet{PouquetETC76} and \citet{MeneguzziPRL81}
. The reason is that, in this situation, while energy, either kinetic or magnetic, is equipartitioned into each helical mode, %whatever the value of cross helicity (or $\beta_C$),
magnetic helicities of both sectors with opposite signs are ``attracted'' by the same pole $k_p=0$%, indicating still a mechanism of dynamo
. %Such attraction is, by simple local eddy turn-over time argument, weaker than the condensation around a finite positive pole as the $\beta_M\ne 0$ case.
Note that unlike EMHD in \S\ref{sec:EMHD}, magnetic energy itself here is not conserved and kinetic energy can be transformed to it to ease the inverse magnetic helicity transfers for the two chiral sectors simultaneously.
Note also that \citet{PouquetETC76} and \citet{MeneguzziPRL81}, and other isotropic MHD simulations with unit Prandtl number, found a slight excess of magnetic energy at small scales, which may be due to such ``attraction'' from large scales.
Without decomposition, as net helicity at any $k$ is seen to be zero, researchers traditionally have not seriously thought about the simultaneous backward transfer of both sectors, to our best knowledge.

\subsubsection{Two-fluid MHD}\label{sec:tfMHD}
Two-fluid effects, the decoupling between electrons and ions, are important in many laboratory %, such as magnetic fusion facilities,
and astrophysical situations \citep[see, {\it e.g.},][]{YamadaETC02,BrandenburgSubramanian05}. % but not well understood.  %Without ``thermal corrections'' (viscosity, resistivity and mutual drag), t
The ideal incompressible two-fluid MHD states that the generalized vorticities $\nabla\times\bm{P}_s$, with canonical momenta $\bm{P}_s=m_s\bm{u}_s+q_s\bm{A}$ for each species $s$, are ``frozen in'' ({\it c.f.} \S\ref{sec:EMHD}) to the respective flows \citep[see, {\it e.g.},][and references therein]{VictorPRuban99}
%\begin{eqnarray}
%&&\nabla  \cdot {\bm{u}}_s= 0,   \label{eq:incomp} \\
%&&
$${m_s}{n_s}\frac{{d{{\bm{u}}_s}}}{{dt}} = {q_s}{n_s}(\bm{E} + {{\bm{u}}_s} \times \bm{B}) - \nabla {p_s},$$ %, \hfill \label{eq:Euler} %\\
%&&\frac{{\partial \bm{B}}}{{\partial t}} + \nabla  \times \bm{E} = 0, \hfill \label{eq:Faraday}\\
%&&\frac{{\partial \bm{E}}}{{\partial t}} - \nabla  \times \bm{B} =  -\sum\limits_s {{q_s}{n_s}{{\bm{u}}_s}} \hfill   \label{eq:Ampere}\\
%&&\nabla  \cdot \bm{B} = 0 \hfill   \label{eq:GaussB}\\
%&&\nabla  \cdot \bm{E} = \sum\limits_s {{q_s}{n_s}} ,        \label{eq:GaussE}
%\end{eqnarray}
where $\bm{E}$ is the electric field vector and $q_s$ and $m_s$ are charge and mass.
Since this model has very rich physics, to remind ourselves the relevant context and the weights of physical quantities in quantifying chirality, instead of being purely geometrical as discussed in \citet[][]{Petitjean03}, we now use the normalization which keeps some physical parameters explicitly, unlike those in EMHD and single-fluid MHD.
The dynamics is constrained by three rugged invariants, {\it i.e.}, the total energy and self-helicities:
$$\mathcal{E} = \frac{1}{2\mathcal{V}}\int \big{[} \bm{E}^2 + \bm{B}^2+\sum_{\tilde{s}} m_{\tilde{s}} n_{\tilde{s}}\bm{u}_{\tilde{s}}^2 \big{]} d^3\bm{r} \ \text{and} \ {\mathcal{H}_s} = \frac{1}{{2\mathcal{V}}}\int \nabla\times\bm{P}_s \cdot \bm{P}_s d^3\bm{r}.$$
Here, two-fluid effects are in the extra terms in the invariants compared to single-fluid MHD.

With Eq. (\ref{eq:FourierHelical}), we are led to the following AE spectra densities ({\it c.f.}, \S\ref{sec:hae}, also for notations)
\begin{eqnarray}\label{eq:tfmhdsd}
Q_M^c(k)&=&\frac{c}{k}U_M^c(k), \
U_M^c(k)
 = \frac { kT(k) }{
D^c(k)
}, \ %&&\\
%&&\\
%\text{and} \ Q_{sK}^c(k)=c k U_{sK}^c(k), \ %&&\\
%Q_{sC}^c(k)
%   = \frac{
%%   c m_s N_s^c k
%%-2m_s q_s^2
%-q_s\beta_s(\alpha n_{\bar{s}}+c\beta_{\bar{s}}m_{\bar{s}} k )k
%%+c q_s^2 \prod\limits_{\tilde{s}}\left( \alpha\,n_{\tilde{s}}+c\beta_{\tilde{s}}m_{\tilde{s}}k \right)
%   }{D^c}. &&
U_{sK}^c(k) =  \frac { N_s^c(k) } { D^c(k) }, \
Q_{sC}^c(k)
   = \frac{ q_s\beta_s L_{\bar{s}}(k)
%   c m_s N_s^c k
%-2m_s q_s^2
%(\alpha n_{\bar{s}}+c\beta_{\bar{s}}m_{\bar{s}} k )
%+c q_s^2 \prod\limits_{\tilde{s}}\left( \alpha\,n_{\tilde{s}}+c\beta_{\tilde{s}}m_{\tilde{s}}k \right)
   }{-D^c(k)/k},\nonumber\\
Q_{sK}^c(k)&=&c k U_{sK}^c(k)\ \text{and} \ %&&\\% \ \text{with}
Q_s^c(k)=m_s^2Q_{sK}^c(k)+2m_sq_sQ_{sC}^c(k)+q_s^2Q_M^c(k)
\end{eqnarray}
with $
D^c(k)=\alpha  k [T(k) +%\beta_{{e}}\,\beta_{{i}}
%( {q_{{e}}}^{2}n_{{e}}m_{{i}}+ {q_{{i}}}^{2}n_{{i}}m_{{e}} )
(\small{\sum}_{\tilde{s}}q_{\tilde{s}}^2n_{\tilde{s}}m_{\bar{\tilde{s}}})\small{\prod}_{\tilde{s}}\beta_{\tilde{s}}]+\underline{c\alpha O\small{\prod}_{\tilde{s}}n_{\tilde{s}}%n_e n_i O
}$, $L_s(k)=\alpha\,n_{{s}}+c\beta_{{s}}m_{{s}}k$,
%\!\!\!\!\!\!\!\!\!\!\!\!
$m_s N_s^c(k)=k [\alpha L_{\bar{s}}(k) + q_{s}^{2} m_{\bar{s}}\small{\prod}_{\tilde{s}}\beta_{\tilde{s}}%\beta_e\beta_i
]+\underline{c n_{\bar{s}} O}$, $O= \alpha\small{\small{\sum}}_{\tilde{s}}q_{\tilde{s}}^{2}\beta_{\tilde{s}}%q_e^{2}\beta_e+q_i^{2}\beta_i
$ and $T(k)=\small{\small{\prod}}_{\tilde{s}}L_{\tilde{s}}(k)%L_e(k) L_i(k)
$
where $\bar{s}$ means the other species than $s$ and where the index $C$ is for the ``cross'' helicity as defined in \S\ref{sec:sfMHD}.
We summarize the following points:\footnote{The result unavoidably appearing a bit complicated, it may be helpful for readers to focus on $U_M^c(k)$ and $Q_M^c(k)$ first. Further simplification of these formulae can be made for some situations.
For instance, electron-positron plasma with mass equivalence and charge conjugation enables us to take all masses and charges be normalized unity.
But, for our purposes here, taking
$D^c$ as polynomials of $k$ with the fundamental theorem of algebra and Vieta's formulas in mind suffices% for our desired level of understanding the behaviors of the spectra
. Note that we have general formula for the roots whose nature is determined by the discriminant. In the cases discussed below, the parameter regimes (constrained by the realizability condition) can be obtained with such basic knowledge by some simple but tedious manipulations and are omitted here. Electric energy distribution is omitted for two reasons: One is that it can be neglected in usual cases; and, the second is that it is decoupled from the others.
}

\textit{First}. The poles, {\it i.e.},
%(nonzero --- see the {\it Last} item below)
roots of the third order polynomials $D^c(k)$, of the two sectors are of opposite signs, as $c$ appears in the second and zeroth order terms.

\textit{Second}. The relevant spectrum may be of {\huge $\llcorner$} shape, with a positive pole on the left. This is similar to the single-fluid MHD (\S\ref{sec:sfMHD}) case with OCSDS of inverse magnetic helicity transfer. Like the discussion in the end of \S\ref{sec:EMHD} for the regime of scales much smaller than $d_e$, {\huge $\lrcorner$} shape spectral density as in HD (\S\ref{sec:HD}) may be relevant to the scenario of both energy and helicities cascading forwardly. Now there can be other larger positive poles, from the same chiral sector or not, which may also be physically relevant to the possible persistence or emergence of chiralities due to the dominance of different physical processes at different scale regimes. Note that like Hall MHD in \citet{MeyrandGaltierPRL12}, ion MHD (IMHD) or EMHD
%(with finite $d_e$, unlike the reduction from Hall MHD)
can be identified in our two-fluid model by putting the fluid velocity of one species to zero: Asymmetries between the respective dynamics, with the opposite chirality in their sense of whistler or cyclotron waves, may present at different scale regimes, due to, say, the different $m_s$ of the two species (unlike electron-positron plasma).\footnote{The chirality of \citet{MeyrandGaltierPRL12} designated by their magnetic polarization does not depend on our sense of chiral sectors in the IMHD or EMHD linear dispersion relations exploited by them, but only on the signs of the charges. In our work, chirality refers to the definite right- or left-handed sector in the helical representation, which is more basic and works for any models, and which may be used to similarly interpret strong turbulence (not limited to the linear wave dispersion relation) as well.% Also, one can isolate one chiral sector from the helical decomposition without resorting to the linear dispersion relation.
}
%Further looking into the fluxes as \citet{CCE03} may see more information about the chiral symmetry breaking studied there. Note that the {\it last} item below may correspond to their simulation with zero magnetic helicity injection, and in principle similar analysis as in \S\ref{sec:HD} for the case of \citet{CCE03} can be performed.

\textit{Third}. A $\bigsqcup$ shape spectral density [$Q_M(k)$, say] may be confined inbetween two distinct positive poles, belonging to the same chiral sector or not,
which presents cross-scale, {\it e.g.}, going across the ion or electron skin depthes etc., behavior: The general EMHD in \S\ref{sec:EMHD} with finite electron inertial can already have such a feature as is clearly seen from $U_M^c(k)$ given there. The general effects can be understood with the combination of the two cases in the above {\it second} item. And,
the large-$k$ peak may also be relevant to
small-scale field generation \citep[see, {\it{e.g.}}][in the context of battery mechanisms]{BrandenburgSubramanian05} or indicating the ``second order'' or ``implicit'' OCSDS as discussed in \S\ref{sec:HD}
.%[may be relevant to the battery mechanisms: see, {\it e.g.}, \citet{BrandenburgSubramanian05}.]

\textit{Last}. If $O=0$ in the zeroth order terms underlined below Eq. (\ref{eq:tfmhdsd}), magnetic helicity does not act in constraining the AE ensemble through $\sum_{\tilde{s}}\beta_{\tilde{s}}\mathcal{H}_{\tilde{s}}$ ({\it c.f.}, \S\ref{sec:hae}). % is then absent
Note that $k=0$ does not become a pole for this situation, as every spectrum in Eq. (\ref{eq:tfmhdsd}) has a factor of $k$ to cancel it.
Since now $%\beta_e\beta_i
\small{\prod}_{\tilde{s}}\beta_{\tilde{s}}<0$, %as partly indicated in Eq. (\ref{eq:tfmhdsd})
we can see that this case is similar to the last situation considered by \citet[][p. 1423]{k67}, with our $[T(k)]^{-1}>0$ from the realizability conditions being eligible to act the role of $k^2$ there; see also $\bar{U}_K(k)$ in \S\ref{sec:hae}. This corresponds to a {\huge $\lrcorner$} shape spectral density.
The pole for large-scale concentration is gone now, since only $Q_M^c$, by definition, has $k=0$ as the asymptotic pole as in single-fluid MHD. \textit{We thus can infer from this point that magnetic helicity constraint under two-fluid framework is still crucial for large-scale concentration of magnetic fields, as in single-fluid MHD where it is conserved.}

Careful analyses with appropriate choices of the physical parameters can be made for detailed illustration and more subtle implications.
Similarly is for other intermediate models such as Hall MHD and general EMHD with finite electron inertial in \S\ref{sec:EMHD} %which is also Hamiltonian with the canonical momenta (see, e.g., \cite{VictorPRuban99,Eyink09}) $\bm{P}_i=m_e\bm{u}+q_i\bm{A}$ and $\bm{P}_e=q_e\bm{A}$ (resulting in a mixed helicity \cite{Turner86}), from which one can perform similar calculation and analysis. The
whose result is a bit simpler, with the denominators being polynomials of second order. From the plasma physics point of view, it is very interesting to spell out all detailed effects of each physical element (skin depths, mass ratio effects etc.), which however is not the focus of this paper. %and would be communicated elsewhere with also the help of other tools \cite{ZouETC}.
We have to refrain from treating these cases in too great a detail.
%Residuals of AE results may persist but require the assistance of other tools for more accurate analyses.

%The major contribution of this subsection is confirming and further illustrating three things: One is the generic OCSDS, due to approaching the pole of one chiral sector of AE spectrum; the importance of magnetic helicity; ``kinetic'' effects as represented by small-scale two-fluid statistics (see the typical discussion in the {\it third} item).

\section{Further discussions %Conclusive discussions
}\label{sec:Conclusion}

\subsection{General remarks %Conclusion
}\label{sec:GR}
%Just as the (Gauss) linking number (the imbalance of the positive and negative crossings) is a rather rough invariant of linkage, so is helicity
%%(whichever one we have studied here)
%for turbulent fields.
%Looking into the finer structures has proved to be helpful.
Turbulence statistics %, such as those of \citet{k73} and \citet{FrischETC75} (including other dynamo discussions of this latter school),
can be sharpened with helical representation, which has been well discerned since \citet{Moffatt70} and K73 \citep{k73} and has become fully workable since \citet[][]{Moses71} and \citet{W92}. %Not much has been done.
We have merely focused on the most basic AE aspect concerning the direction of spectral transfer as well as the selection and amplification of chirality.
%, which can be quantified by helicity and relevant derivatives such as the relative helicity \citep{k73}. %A lot more work can be done.
The key point is that although the dynamics of the two chiral sectors are in general coupled, the absolute equilibrium spectra are cleanly split, with poles of opposite signs. Not only that the finer physical structures offer new insights about the ``(near-)racemic mixture'' ({\it c.f.}, the last paragraph of \S\ref{sec:sfMHD}), but also that one should keep away from the misconception that the chirally decomposed quantities derived in this paper never
appear in the AE ensemble by themselves, but always in combinations giving inviscid invariants. Actually, OCSDSs may emerge in natural systems due to mechanisms relevant to what we have discussed or one can work with samples of ``enantiopure compounds'' in \S\ref{sec:HD}. Concerning partial fraction decomposition, our results physically assures the decomposability of the spectra of the traditional Fourier modes from the hydrodynamic-type models studied here and practically solves the mathematical problem, giving also the nice ``conjugate'' mathematical structures in the spectra of the opposite chiral sectors, at least to the degree of two parts with poles of exactly opposite signs, which is not trivial for some models such as the two-fluid MHD in \S\ref{sec:tfMHD}. Naive attempts to perform the ``post'' decomposition of the traditional spectrum could be formidable and confusing, for lack of physical motivation: For instance, one could think of trying further to decompose the already chirally decomposed two-fluid MHD spectra.%into two independent sectors.
%And the pole(s) of AE spectra of opposite chiral sector are of opposite sign, which makes one of the sector overwhelmingly dominate around its pole.
%Some reported relevant numerical results have been accordingly addressed/explained, but evidently much more numerical work carefully looking into the two chiral sectors are still wanted.

For the absolute equilibria themselves, both K73 and \citet{FrischETC75}'s insights were almost here as we are now. Especially K73 explicitly discussed the interactions of pure helical modes. Curiously, K73 however did not\footnote{\citet{k73} might not have been motivated to systematically examine the dynamics in helical-mode representation, especially the Liouville theorem and detailed conservation laws, the latter of which only appeared two decades later \citep{W92}. There is a conceptual issue here as we will elaborate a little bit. He wrote in the second page of that paper: ``The two helical waves provide an alternative to the usual Fourier decomposition into plane-wave components.'' In the usual Fourier representation, with consideration of isotropy (but lack of reflexion symmetry) as he was considering, there is no reason and no way to distinguish special components of the 3D spectra or spectra of special components of the Fourier coefficients, since they are all statistically identical. And, now the helical representation, as he noted, is only an alternative to the usual Fourier representation, concerning the degrees of freedom, thus he might omit the important distinguishability of the spectra between the opposite sectors.} study the chiroids absolute equilibria to which his traditional mixed ones can not be reduced by taking any limits of any of the temperature parameters, in which sense we mean, in \S\ref{sec:HD}, his results are not valid for asymmetrically truncated systems. Such a piece of thin ``window paper'' was not pierced probably due to the fact that conventionally the relevance of the traditional chirally symmetric Galerkin truncation were made to the classical chirally symmetric viscosity or resistivity following \citet[][footnote 2]{Lee1952} and \citet[][footnote 8]{k73}. %\footnote{
For such physical considerations, see also some recent works \citep{FrischETCprl08,ZhuTaylorCPL10} where other dissipation models lead to convergence to the classical Galerkin truncations in some sense. %}
However, with our HD in \S\ref{sec:HD} results, we have refined Kraichnan's argument to reveal an interesting feature of helical turbulence beyond the non-existence of negative temperature states that Kraichnan emphasized. Just as Kraichnan used the AE to suggest directions of energy transfer, so that a pile-up of energy at large scales in the 2D case can suggest the flux of energy to large scales far from equilibrium, we suggest that the preferential transfer into one chiral sector observed by \citet{CCE03} might be related to our observation about his helical AE. %\textbf{Probably more surprising or exciting is that, for controlling roughness of the solutions to transit ``smoothly'' from inverse to forward energy cascades, we have proposed a truncation scheme which was motivated by pure academic quest but which turns out to be relevant to rotating turbulence.}
Furthermore, if we can somehow introduce chirally asymmetric dissipation and/or resistance in (magneto)fluids\footnote{It might be possible to work with some chiral (conducting) polymers; or, for classical magnetofluids, some special electromagnetic techniques would be wanted. Note that conventional study of elastic polymer effects \citep[such as][]{PLBrmp08,SteinbergPhysique09} have not paid attention to the chirality, that is, the possibility of a third chiral time scale $\tau_{\theta}^c$, over which the (chiral) torque is to be balanced, besides the transverse and longitudinal ones \citep[ $\tau_{\perp}$ and $\tau_{\parallel}$ in][]{HatfieldQuakePRL99} of the extended coil/helix (such as DNA), and that in a simple dilute polymer solutions dynamical model \citep[][whose nonlinear dynamics is exactly the same as the classical 3D single fluid MHD studied in \S\ref{sec:sfMHD}!]{FouxonLebedevPoF03} only a single relaxation time $\tau$ is used for all modes of $\bm{B}$, the so-called ``tau approximation''. It is possible that the 3D chiral property of the polymers have non-neglectable rheological effects, especially in the turbulent states where small-scale helical modes are excited.}, then the small-scale damping as discussed in \S\ref{sec:HD} would not be simply only for chiral symmetry restoration and that explicit small-scale chirality selection and amplification similar to those at large scales could also present.
%And the physical relevance of our result could be more explicit.
Actually, in general plasma dynamics, such as cyclotron damping, essentially a 3D analog of the classical 1D Landau damping, and plasma heating \citep[see, {\it e.g.}, Chaps. 10, 11 and 17 of][]{StixBook92}, our result may be of stronger connection, since the ion and electron cyclotron resonances are of opposite chiralities and at different scales, addressable by two-fluid MHD model in \S\ref{sec:tfMHD}.

\subsection{Comparisons: for genericities, specificities and beyond}
%\subsubsection{Magnetic versus kinetic helicities}
A major purpose of this work is to find the genericities and differences of various helicity effects by comparisons of the different hydrodynamic-type models.
The subject in the center of the comparison is that of magnetic versus kinetic helicities.
The pure magnetodynamic result for OCSDS of magnetic helicity (transfer) at large scales, as represented in the vanishing-$d_e$ EMHD, generically lies in the core also of other MHD models. %Such large scales appears to be close to equilibrium (Gaussian).
Two-fluid MHD has the most general and complete elements of helicities and show convincingly the crucial role of magnetic helicity for large scales. It appears to be nothing deep but simply due to the mathematical relations $$Q_K^c(k)= kU_K^c(k)$$ and $$Q_M^c(k)= U_M^c(k)/k$$ by definitions, by which, one can practically assume equipartition between kinetic and magnetic energy of same chiral sector at some intermediate scale and find that magnetic helicity of that sector belongs more to larger scales. One ``artificial'' way to look at it is the following: The gyrofrequency of a charged particle's helical motion around $\bm{B}$ is $\Omega=qB/m$, which means that we can formally take the $\bm{B}$-line as ``kinetic vorticity'' $\bm{\Omega}$-line, macroscopically;
%\footnote{
magnetic field is indeed a \textit{pseudovector}, like fluid vorticity, allowing the well-known dynamical analogy between them as initiated by Batchelor \citep[see, {\it e.g.},][]{Moffatt08}. This then gives various helicities a kind of unified description% as the self-helicity of two-fluid MHD
. Magnetic helicity is thus related to the more ``intrinsic'' plasma particle motion. % like quantum spin, and conservation of self-helicities is reminiscent of the orbital and spin angular momenta, and, the conservation of the total. %, besides the analogy between one-chiral-sector HD and 2D gyrokinetics \citep{Zhu12} as we mentioned in \S\ref{sec:HD}.
There are also other supports of the robustness of magnetic helicity \citep[see, e.g.,][]{BrandenburgSubramanian05}, such as analysis with more general context \citep[][]{BergerFieldJFM84} and measurements \citep[][]{JiPRL99}.

The HD and the large-$d_e$ EMHD situations are different to the others, in the sense that the realizable AE spectra can only have positive pole at large $k$s which regime however is subject to dissipation in real physical systems. That is, reflexion symmetry breaking and restoration mechanisms meet at the same battlefield and they reach another kind of equilibrium balance which is far from our statistical absolute equilibrium, whose implications and residuals concerning chirality in conventional fluids can still be identified with careful analyses as shown in \S\ref{sec:HD} by refining Kraichnan's argument.
Restricting to the homochiral situations \citep{W92,bmt12}, HD kinetic energy or EMHD magnetic helicity accumulating at or transferring to large scales becomes possible as indicated by the negative-$\alpha$ state with small-$k$ pole, and we further find that, as long as the asymmetry is strong enough, adding aliens to exclude the negative temperature state does not necessarily drastically change the transfer picture. It is possible to control the smooth transition from completely inverse to partly inverse and partly forward and to completely forward transfers.

Concerning turbulence cascade in the infinite Reynolds number limit or some kind of thermodynamic limit in the sense of $k_{max} \to \infty$ in the conventional Galerkin truncation, full 3D Navier-Stokes' energy and helicity both cascading to small scales indicates that the solution is singular. Note that such an indication however has not yet found rigorous mathematical support and that there is still space for opposite conjectures, such as a solution as some kind of ``directional limit'' without such dissipative anomalies \citep{ZhuTaylorCPL10}. \textit{Careful examination of chiroids absolute equilibria as partly illustrated in \S\ref{sec:HD} turns out to be able to give fine and clear intuitive pictures about the roughness of the solutions.} Now, for homochiral 3D Navier-Stokes with, say, $c=+$, it is expected that only helicity, but not energy, is transferred to small scales, which indicates that the solution is slightly less singular, with the H$\ddot{o}$lder exponent $h$ in $$\delta u(\ell)\sim \ell^h$$ be some value inbetween $1/3$ and $2/3$ \citep{EyinkEulerPhD08}, of course in some statistical sense as the multifractal spectrum of $h$ spans over a wide range in realizations \citep{FrischBook}. The self-similar pure kinetic helicity cascade spectrum would go as \citep{BrissaudETC73,W92}  $$H^+_K(k)\sim k^{-4/3}$$ which, unlike Kraichnan's 2D enstrophy spectrum $\sim k^{-1}$, is convergent when integrated over $k$. Note that now $h=2/3$. This convergence so far does not bring any troubles: Unlike 2D Euler, where finite enstrophy ensures the smoothness of the solution and that in principle ensures an equilibrium statistical mechanics without truncation \citep{Miller90,Robert03}, there is no mathematical theorem to assure conservation of helicity with its finiteness \citep[see, {\it e.g.},][]{EyinkEulerPhD08}. Of course, such cascade still may have spacial intermittency, in the sense of Onsager \citep{EyinkEulerPhD08} that the helicity dissipation appears ``spotty'', in which case an anomalous part of the dissipation may be considered to be \textit{curdling} with infinite density on some fractal sets of zero volumes \citep[][]{MandelbrotJFM74}, described by some Dirac delta function supported by the fractal, as an extremal limit. Coming back to full Navier-Stokes and supposing both energy and helicity cascading forwardly as $k^{-5/3}$, we immediately see from Eqs. (\ref{eq:QU}) and (\ref{eq:QUc}) that $$H^c_K(k)=cC_1k^{-2/3} + C_2k^{-5/3}$$ with $C_1$ and $C_2$ being constants \citep{DitlevsenGiulianiPoF01}. That is, the two sectors of helicity both present ultraviolet divergences, indicating more singular solutions, which is consistent with the absolute equilibrium spectra showing poles at large $k$, compared to the small-$k$ poles for the homochiral case. % and which is, on the other hand, also consistent with the well-known fact that nonlinearity is completely depleted in the homochiral and homowavelength case (see the ending remark in \S\ref{sec:HR} and Eq. \ref{omegag}).
Thus, \textit{as indicated by the smooth transition from backward, to bi-direction and to forward cascades by manipulating the helical modes in \S\ref{sec:HD}}, how the added ``alien'' helical modes increase the nonlinearity to roughen the solutions is intriguing and may be relevant to the intermittency property in the sense of Onsager and Mandelbrot as mentioned above.

\subsection{Conclusion and prospects%: Where can we go from here?
}
Since an incompressible hydrodynamic-type system can be reduced to the dynamics of chiroids {\it a la} Kelvin, it is natural that one reduces the statistical dynamics to what is based on them and ``hopes that one can get some insight into the nature of more general viscous flows and even, perhaps, a deeper understanding of turbulence.'' \citep{Moses71} We have studied the chirality issue, starting from and essentially based on the chiroids absolute equilibria.
The equilibrium spectra can also be used to guide and benchmark numerical experiments with truncation schemes such as those discussed in \S\ref{sec:HD} with asymmetric truncation between the two chiral sectors. Hints for further theoretical considerations may also be inferred. For instance, the clear OCSDS for large-scale magnetic helicity could imply some clues to dynamical dynamo model. Looking further into anisotropic fluctuations with a background magnetic field \citep[for such discussions under the framework of 3D gyrokinetics, see, {\it e.g.},][]{gkaeOLD} and to more realistic laboratory situations is also a reasonable step towards a more comprehensive theory for multi-scale plasma dynamics.
And, due to the cross-disciplinary popularity of the notion of chirality as the legacy of Pasteur and Lord Kelvin \citep[see, {\it e.g.},][]{BarronQJM97}, one may not be able to resist a thought excursion into other fields, such as biochirality
%\footnote{The expectation made in \S\ref{sec:BoTCS} that homochiral ``life'' time would be longer is somewhat motivating, though we don't hastily generalize it.}
\citep[see, {\it e.g.},][]{BlackmondColdSpringHarbor10} among others, which is the reason why we choose the terminology ``chirality'' instead of ``parity'', which may be thought to be associated with the symmetry of fundamental physical laws, or pure geometrical symmetry, {\it i.e.}, mirror symmetry, of objects, or ``polarization'' which is used more for (linear) waves.

%\section{Acknowledgement}
In conclusion, Appendix \ref{app:basicsoftruncation} is written thanks to a referee who raised the questions, believed in their popularity among readers and asked for explicit answers in the paper, and who is also acknowledged for the decompression of the early version of the manuscript.
This work was partially supported by the Fundamental Research Funds for the Central Universities of China and by the WCI Program of the NRF of Korea [WCI 2009-001].
We thank for the discussions with  U. Frisch and S. Kurien on helical hydrodynamic closures and  with M. Taylor on direct numerical simulations of helical absolute equilibria back to half a solar cycle ago, with Z.-B. Guo, Z.-W. Xia and D.-D. Zou on plasma waves, with X.-P. Hu and Z. Lin on plasma heating, and the correspondences with L. Biferale, C.-K. Chan, P. Diamond, D. Escande, M. Faganello, S. Galtier, J. Miller, M. Petitjean and V. P. Ruban, A. Schwartz during the course of this work.
is grateful for the hospitality of the International Institute for Fusion Science, Universit\'e de Provence, and for
the workshop ``The Solar Course, the Chemic Force, and the Speeding Change of Water'' at NORDITA (2011).

\appendix
\section{Very basic aspects of the truncated system}\label{app:basicsoftruncation}

\subsection{On the detailed conservation laws for the pure helical modes in each interacting triad}\label{app:conservation}

Let us outline here, for the HD case, a direct verification \citep[{\it e.g.},][]{W92} and a proof \citep[{\it e.g.},][]{k73} of the detailed conservation laws for energy and helicity of the pure helical modes among each interacting triad. Both of them are simply carried over from those for the traditional Fourier modes.
The direct \textit{verification} starts from the dynamical equation of the pure helical modes, with the interactions restricted among only one triad as given by Eq. (9) of \citet{W92}. As he shows, simple algebras by the definitions of energy and helicity using this equation then verify the conservations of energy and helicity of the three conjugate pairs of pure helical modes, regardless the handedness of any chiroid.
The alternative \textit{proof} also needs only to change the objects of the classical Fourier modes, in the third paragraph in p. 748 of \citet{k73}, to pure helical modes. The idea is simply that the overall energy and helicity are formally conserved by the original dynamics without explicit truncation and the truncated modes' energy and helicity are constantly zero, due to the facts that their amplitudes are set to be nulls by definition of truncation and that the `energy and helicity expressions are quadratic and diagonal in the wave-vector amplitudes.' Note that the expression being diagonal in the wave-vector amplitudes is also important: Suppose it is not diagonal and that the convolution involves the multiplication of modes in the truncated and un-truncated domains, then the change rate of it is not assured to be constantly zero, since the change rate constitutes a component from the multiplication of the time derivative of a mode in the truncated domain with another mode in the un-truncated domain, both of which can be non-zero; an example is the quadratic invariants of 2D gyrokinetics in the Fourier-Hankel/Bessel representation and truncation, where the phase-space ``wave-vector'' is extended from the conventional wave-vector to include a component from the spectral representation of the velocity variable and where those quadratic expressions not being diagonal in the extended wave-vector are not ruggedly conserved, {\it i.e.}, not invariant after Fourier-Hankel/Bessel truncation \citep[see pp. 3--4 of][]{ZhuFourierHankel11}.
%Such proof by Kraichnan is the most elegant, to our best knowledge.

As some readers may feel easier to start with a degenerate trivial case to get somewhat more concrete grasp, let's suppose first that we
retain only contributions from the region of wave-number space $|\bm{k}| < K$ (Galerkin
truncation), and that we start with a single triad of pure helical modes (chiroids), using indexed $c$ to denote the chirality of the leg, $\{[\pm\bm{k},c_{\bm{k}}];[\pm\bm{p},c_{\bm{p}}];[\pm\bm{q},c_{\bm{q}}]\}$ with $\bm{k} + \bm{p} + \bm{q} = 0$
and $K/2 < |\bm{k}|,\ |\bm{p}|,\ |\bm{q}| < K$, so that harmonics generated
by the Euler equations are eliminated under this truncation which we can think of as
providing some kind of `artificial dynamics' and which is denoted by the
%operator $\mathcal{T}$
index ``$_g$''. Let $\bm{u}_g(\bm{x}, t)%=\mathcal{T}\bm{u}(\bm{x},t)
$ and $\bm{\omega}_g(\bm{x}, t) = \nabla \times \bm{u}_g(\bm{x}, t)$ be
the velocity and vorticity fields evolving under this artificial dynamics. Then the claim
is that the mean kinetic energy $\mathcal{E}_{g} =< \bm{u}_g^2/2 >$ and mean helicity $\mathcal{H}_{g} =< \bm{u}_g \cdot \bm{\omega}_g >$ are
invariant in time. Yet another degenerate case is the Arnold-Beltrami-Childress (ABC) flow, composed of three conjugate pairs of pure helical modes with the same wavelength, which has been used by many authors to study kinematic dynamo and which can be generalized to contain more conjugate pairs of pure helical modes of same wavelengths and that presumably to become more chaotic, in the Lagrangian/streamline sense \citep[see, {\it e.g.},][and references therein]{ArnoldKhesin98book}.

\subsection{Dynamical and topological aspects}\label{app:dynamics}
By definition, the Galerkin-truncation dynamics of vorticity is
\begin{equation}\label{omegag}
    \partial_t\bm{\omega}_g=%\mathcal{T}
    [\nabla \times (\bm{u}_g \times \bm{\omega}_g)]_g.
\end{equation}
%[With the very specific Galerkin truncation considered in Appendix \ref{app:conservation}, obviously $\partial_t \bm{\omega}_g=0$, since the right-hand side of Eq. (\ref{omegag}) leaves no modes between $K/2$ and $K$.]
We are not sure whether the fact that $\mathcal{H}_g$ is constant assures a fictitious meaningful $\tilde{\bm{v}}$ to solve $\partial_t\bm{\omega}_g=\nabla \times (\tilde{\bm{v}} \times \bm{\omega}_g)$, {\it i.e.}, making an analogue of the Kelvin-Helmholtz or ``frozen-in'' theorem (which is sufficient but not necessary for the conservation law.) Definitely, $\tilde{\bm{v}}=\bm{u}_g$ is not the solution, otherwise the last index ``$_g$'' on the right-hand side of Eq. (\ref{omegag}) would have no effect. %, though such fictitious velocity $\tilde{\bm{v}}=\bm{u}_g$ does lead to invariance of $\mathcal{H}$.

Note that the topological interpretation of the helicity as the degree of (average) knottedness and/or linkage \citep{Moffatt69} of (closed) field line(s) formally carries over to the Galerkin truncated case. Actually, the interpretation itself has not much to do with the dynamics but simply works for fields satisfying some basic properties by the definition of Gauss linking number, which has a lot of subtleties and complications when generalized to continuous fields and general boundary conditions \citep[see, {\it e.g.},][among many other references cited therein and appearing later]{Moffatt69,BergerFieldJFM84,ArnoldKhesin98book}.
%For simplicity, let's start with the quantized case, that is, the case with the field(s) curdling on the flux filament(s). Let $\nabla\times\bm{F}_1$ and $\bm{F}_2$ be the two divergence-free vector fields, identical or not, and $\textit{C}_2$ be a closed $\bm{F}_2$ filament. Then, $\oint_{\textit{C}_2}\bm{F}_1 \cdot \bm{F}_2 d\ell$
The Galerkin-truncated dynamics is formally changed much in physical space, as partly shown in the last paragraph, while formally unchanged (except for truncation) in Fourier space concerning triadic interactions. One formally unchanged thing, besides those such as the quadratic invariants, that clearly bridges the physical- and Fourier-space representation, is the preservation of the incompressibility of the fields, which is also due to the fact that the orthogonality between their chiroids and the wavevectors holds in detail, {\it i.e.}, for each chiroid $\bm{k}\cdot\hat{\bm{h}}_c(\bm{k})=0$, and which justifies the definition of flux tube(s) as the key to the topological interpretation \citep{Moffatt69}. Topological, especially knot-theory, approaches to the statistical dynamics of course deserves further pursue, which is however beyond the scope of this note.

\subsection{On the tacit assumptions in the statistical consideration}\label{app:assumption}
It is difficult and quite open to establish mathematically rigorous conditions for justifying the application of statistical mechanics. For example, according to literatures \citep[see, {\it e.g.},][and references therein]{EyinkSreenivasan06}, ergodicity, being sufficient, may not be trivially satisfied but may neither be necessary; and, the mixing time scale could be hard to estimate for evaluating the closeness of physical relevance of the equilibrium ensemble.
However there is a trivial bottom line that is assumed to be met, that is, all modes should be directly or indirectly connected by forming the interacting triads to define a system.
For example, now suppose instead that we start with two triads $\{[\pm\bm{k}_j,c_{\bm{k}_j}];[\pm\bm{p}_j,c_{\bm{p}_j}];[\pm\bm{q}_j,c_{\bm{q}_j}]\}$ with $j=1,2$ and
all these wave-vectors in the spherical annulus $(K/2, K)$, as in Appendix \ref{app:conservation}, and suppose that
these triads are non-interacting. Let the
energies in the triads be $E_{g1}$ and $E_{g2}$ respectively, and the helicities $H_{g1}$ and $H_{g2}$. Then it is not eligible to use the ensemble defined by the total
energy $E_g = E_{g1} +E_{g2}$ and the total helicity $H_g = H_{g1} +H_{g2}$ for the union of these two isolated systems, not to mention that sufficient number of modes are necessary for a statistical consideration, and in particular the application of Gibbs ensemble.
%\footnote{Alternatively, from E. T. Jaynes' statistical inference point of view, we need to take this extra piece of information, that the two triads are not interacting, into account, which neither support the direct naive application of the Gibbs ensemble.}
Actually, in practice, when the number modes is large it is hardly possible for any triad to be isolated from others. In performing the calculations as in the main text, such tacit assumptions are made to exclude cases not describable by the canonical ensemble.
In the 1970s, people already performed many numerical simulations, mostly for 2D cases \citep{Orszag, KraichnanMontgomery}, to study the ergodicity and mixing properties, to measure the difference between microcanonical and canonical ensembles, to find how many modes would be needed to reach the Gibbs state and to finally verify the corresponding energy spectrum. Matthaeus and collaborators, among others, have many followup studies, especially for 2D and 3D MHD absolute equilibrium ensembles. We cannot exhaust the list of references here, but would like to remark that there can be new findings by revisiting numerical check of the tacit assumptions relevant to the application of Gibbs ensembles with the new special truncations on the new freedoms of chiroids.

% susie put cite commands here, don't bother with citet etc just yet.

\bibliographystyle{jfm}

\bibliography{MS4-R}

\end{document}